\documentclass{PoS}

\usepackage{epsfig}
\usepackage{amsmath}
\usepackage{amssymb}
\newcommand\epjc[3]{{\it Eur. Phys. J.}{\bf C #1} (#2) #3}  

\def\be{\begin{equation}}
\def\ee{\end{equation}}
\def\bea{\begin{eqnarray}}
\def\eea{\end{eqnarray}}
\def\bsp{\begin{split}}
\def\esp{\end{split}}

\title{Saturation effects in QCD from linear transport equation.}

\ShortTitle{Saturation effects in QCD from linear transport equation.}

\author{Krzysztof Kutak\\
        Universiteit Antwerpen, Groenenborgerlaan 171, Antwerpen, Belgium\\
        E-mail: \email{Krzysztof.Kutak@ua.ac.be}}

\abstract{We show that the GBW saturation model provides an exact solution to the one-dimensional linear transport equation. We also show that 
it is motivated
by the BK equation considered in the saturated regime when the diffusion and the splitting term in the diffusive 
approximation are balanced by the nonlinear term.}

\FullConference{XVIII International Workshop on Deep-Inelastic Scattering and Related Subjects, DIS 2010\\
		April 19-23, 2010\\
		Firenze, Italy}

\begin{document}
\section{Introduction}
Perturbative Quantum Chromodynamics (pQCD) at high energies can be formulated in
coordinate space in the dipole picture \cite{Mueller}. If we in particular focus on Deep Inelastic Scattering the scattering process
can be described in this picture as interaction of virtual photon which has just enough energy to dissociate into a 'color
dipole' with the hadronic target carrying most of the total energy. The interaction process is described here by the  
dipole-nucleus scattering amplitude. 
This amplitude can be modeled and we will focus here on  
the Golec-Biernat W\"usthoff saturation model \cite{GBW} which 
includes saturation effects. It was motivated by requirements that at the high energy limit of QCD the total cross section for hadronic processes
should obey unitarity requirements. At present there are much more sophisticated approaches to introduce these requirements
in a description of scatterings at high energies \cite{JalilianMarian:1997jx,JalilianMarian:1997gr,JalilianMarian:1997dw,Bal,Kov,Avsar:2006jy,BartKut}. 
However, one can still ask the question if there is any dynamics 
behind the GBW model or to put it differently is there any equation to which formula proposed by Golec-Biernat and W\"usthoff is a solution? 
And what is the role of the initial conditions? In this article we report on answer to these questions provided in \cite{Kutak:2009zk}.\\
\section{GBW model and a transport equation}
The GBW amplitude following form GBW cross section and related to it by $\sigma(x,r)=2\int d^2b N(x,r,b)$ reads (here we are interested in the original 
formulation):
\be
N(x,r,b)=\theta(b_0-b)\Bigg[1-\exp\left(-\frac{r^2}{4R_0^2}\right)\Bigg]
\label{eq:GBW98}
\ee  
where $b$ is the impact parameter of the collision defined as distance between center of the proton with radius $b_0$ and center of a dipole 
scattering on it, 
$r$ is a transversal size of the dipole, $x$ is the Bjorken variable, 
$R_0(x)\!=\!\frac{1}{Q_0}\left(\frac{x}{x_0}\right)^{\lambda/2}$ 
is the so called saturation radius and its inverse defines saturation scale, $Q_s(x)\!=\!1/R_0(x)$ and $x_0$, $\lambda$ are free parameters. 
This amplitude saturates for large dipoles $r\!\gg\!2R_0$ and exhibits geometrical scaling which has been confirmed by data \cite{Stasto:2000er}.
\subsection{Transport equation for unintegrated gluon density}
The dipole amplitude (\ref{eq:GBW98}) can be related to the unintegrated gluon density  which convoluted with the $k_T$ dependent 
off-shell matrix elements allows to calculate observables in the high energy limit of QCD. This relation is the following (after assumption that the dipole is 
much smaller than the target):
\be
f(x,k^2,b)=\frac{N_c}{4\alpha_s\pi^2}k^4\nabla^2_k\int \frac{d^2{\bold r}}{2\pi}\exp(-i{\bold k}\cdot{\bold r})\frac{N(x,r,b)}{r^2}
\label{eq:ugd}
\ee
where $r$ and $k$ are two-dimensional vectors in transversal plane of the collision and $r\equiv|\bold r|$, $k\equiv|\bold k|$\\ 
Performing this transformation we obtain the known result \cite{GBF}:
\be
f(x,k^2,b)=\frac{N_c}{2\pi^2\alpha_s}\theta(b_0-b)R_0^2(x)k^4\exp\left[-R_0^2(x)k^2\right]
\label{eq:ugd2}
\ee
 Now motivated by the fact that function $f(x,k^2,b)$ exhibits a maximum both as a function of $x$ for fixed $k^2$ and as a function
of $k^2$ for fixed $x$, we differentiate $f(x,k^2,b)$ with respect to $x$ and $f(x,k^2,b)/k^2$ with respect to $k^2$.
We obtain:
\be
\partial_x f(x,k^2,b)=\frac{\lambda f(x,k^2,b)(1-R_0^2(x)k^2)}{x Q_0^2}
\label{eq:poch1}
\ee
\be
\partial_{k^2} \frac{f(x,k^2,b)}{k^2}=\frac{f(x,k^2,b)(1-R_0^2(x)k^2)}{k^4Q_0^2}
\label{eq:poch2}
\ee
Dividing eqn. (\ref{eq:poch1}) by (\ref{eq:poch2}) and rearranging the terms and defining ${\cal F}(x,k^2,b)=f(x,k^2,b)/k^2$, $Y=\ln x_0/x$, $L=\ln k^2/Q_0^2$
we obtain:
\be
\partial_Y {\cal F}(Y,L,b)+\lambda\partial_L {\cal F}(Y,L,b)=0
\label{eq:trans}
\ee
which is the first order linear wave equation also known as the transport equation. As it is linear it cannot generate saturation dynamically but it can propagate well the initial condition leading to a successful phenomenology \cite{GBW}. 
It describes the change (wave) in the particle distribution flowing into and out of the phase space volume with velocity $\lambda$. 
This wave propagates in one direction. The quantity ${\cal F}(x,k^2,b)$ gains here the interpretation of a number density of gluons with momentum 
fraction $x$ with the transversal momentum $k^2$ at distance $b$ from the center of the proton. 
The general solution of (\ref{eq:trans}) can be found by the method of characteristics and is given by:
\be
{\cal F}(Y,L,b)={\cal F}_0(L-\lambda Y,b)
\label{eq:sol}
\ee
One can go back from (\ref{eq:trans}) to (\ref{eq:ugd2}) using following initial condition at $x=x_0$: 
\be
{\cal F}(x\!=\!x_0,k^2,b)=
\frac{N_c}{2\pi^2\alpha_s}\theta(b_0-b)k^2\exp(-k^2)\\
\label{eq:gbwini}
\ee
This initial condition has saturation built in, since the gluon density vanishes for small $k^2$.
Knowing the properties of the linear first order partial differential equation we see that the property of saturation
of GBW was a consequence of the wave solution which relates $x$ and $k^2$ supplemented by initial conditions with saturation built in. 
We also see that the {\it critical line} of the GBW saturation model visualizing, the dependence of the saturation scale on $x$, 
$Q_s(x)=Q_0\left(\frac{x_0}{x}\right)^{\lambda/2}$ is in fact from the mathematical point of view the characteristics of the 
transport equation.
\subsection{Transport equation for the dipole amplitude in momentum space}
Similar investigations can be repeated for the momentum space representation of the dipole amplitude  $N(x,r,b)$ 
which we denote by $\phi(x,k^2,b)$. 
\be
\phi(x,k^2,b)=\int\frac{d^2{\bold r}}{2\pi}\exp(-i{\bold k}\cdot{\bold r})\frac{N(x,r,b)}{r^2}
\label{eq:ugd}
\ee
A nonlinear pQCD evolution equation like the Balitksy-Kovchegov (BK) equation written for $\phi$ (in large target approximation) takes quite simple form and 
can be related directly to the statistical formulation of the high energy limit of QCD (see \cite{SM2} and references therein). 
Applying this transformation to (\ref{eq:GBW98})  we proceed with differentiation similarly as before and we obtain:
\be
\partial_Y\phi(Y,L,b)+\lambda \partial_L \phi(Y,L,b)=0
\ee
which is, as before, the transport equation.
\subsection{Relation to pQCD}
It is tempting to investigate the relation between found transport equation and the high energy pQCD evolution equations  like \cite{Bal,Kov,BartKut}. 
Let us focus here in particular on the form of the BK equation in 
large cylindrical target approximation for the dipole amplitude in momentum space for which the nonlinear term is just a simple local quadratic expression.
The BK equation for the dipole amplitude in the momentum space reads:
\be
\partial_Y\phi(Y,k^2,b)=\overline\alpha\chi\left(-\frac{\partial}{\partial_{\log k^2}}\right)\phi(Y,k^2,b)-\overline\alpha\phi^2(Y,k^2,b)
\label{eq:BKdip}
\ee
where $\overline\alpha=\frac{N_c\alpha_s}{2\pi}$  and $\chi(\gamma)=2\psi(1)-\psi(\gamma)-\psi(1-\gamma)$ is the characteristic function of the BFKL 
kernel which allows for emission of dipoles and therefore drives the rise of the amplitude. The role of the nonlinear term is 
roughly to allow for multiple scatterings of dipoles which contributes with negative sign and slows down the rise of the amplitude. This equation 
provides unitarization of the dipole amplitude \cite{BKphen} for fixed impact parameter and admits traveling wave solution in the diffusion approximation\cite{SM1}.
The analytic solution of (\ref{eq:BKdip}) within the diffusion approximation relying on expanding the kernel of (\ref{eq:BKdip}) up to second 
order and mapping it to the Fisher-Kolmogorov equation has been obtained by Munier and Peschanski \cite{MP3}. It reads:
\be
\phi(Y,k^2,b)=\theta(b_0-b)\sqrt{\frac{2}{\overline\alpha\chi''(\gamma_c)}}
\ln\left(\frac{k^2}{Q_s^2(Y)}\right) 
\left(\frac{k^2}{Q_s^2(Y)}\right)^{\gamma_c-1}
\exp\left[-\frac{1}{2\overline\alpha\chi''(\gamma_c)Y}\ln^2\left(\frac{k^2}{Q_s^2(Y)}\right) 
\right]
\label{eq:solBK1}
\ee
where $\gamma_c=0.373$ and $Q_s^2(Y)$ is emergent
saturation scale given by:
\be
Q_s^2(Y)=Q_0^2e^{-\bar\alpha\chi'(\gamma_c) Y-\frac{3}{2\gamma_c}\log Y
-\frac{3}{(1-\gamma_c)^2}
\sqrt{\frac{2\pi}{\bar\alpha\chi^{\prime\prime}(\gamma_c)}}\frac{1}{\sqrt{Y}}
+{\cal O}(1/Y)}\ 
\label{eq:satscal}
\ee
By inspection we see that (\ref{eq:solBK1}) does not obey the transport equation. The problem is caused by the diffusion term.  
However, we can consider the asymptotic regime called "front interior"  \cite{MP3}, region where transverse momenta $k$ is close to the saturation scale 
$Q_s(Y)$ and rapidity $Y$ is large and where the condition $\ln^2\left(\frac{k^2}{Q_s^2(Y)}\right)/2\overline\alpha\chi''(\gamma_c)Y\!<\!\!\!<\!1$ is satisfied.
In this regime (\ref{eq:solBK1}) simplifies and after taking derivatives as in the previous sections we obtain the following wave equation:
\be
\partial_Y {\phi}(Y,L,b)+\lambda_{BK}\partial_L {\phi}(Y,L,b)=0
\label{eq:transBK}
\ee 
 where $\lambda_{BK}=\partial \log Q_s^2(Y)/\partial Y$. In the limit where $\lambda_{BK}$ does not depends on energy \cite{IJM} we obtain:
\be
\lambda_{BK}=-\overline\alpha\chi'(\gamma_c)
\label{eq:asymlimit}
\ee

\section{Conclusions}
In this  note we have shown that the GBW saturation model is the exact solution of a one-dimensional linear transport equation of the form (\ref{eq:trans}).
We conclude that since (\ref{eq:trans}) is a linear equation the saturation property has to be provided in the initial condition.
We found that for the GBW model this equation is universal for the unintegrated gluon density $f(x,k^2,b)$ and the dipole amplitude in momentum space 
$\phi(x,k^2,b)$ but the details of the shape of the wave depends on the initial condition which is different for each of them. We also studied the 
relation of the transport equation to the BK equation in the diffusion approximation. We have shown that in the 
region of phase space where diffusion and splitting processes  are of the same  order as the nonlinear term, the GBW model is consistent with the 
BK equation.


\begin{thebibliography}{99}
\bibitem{Mueller} A.H. Mueller, B. Patel Nucl.Phys. B425 (1994) 471-488
\bibitem{GBW} K. Golec-Biernat and  M. W\"usthoff,
            {\it Phys. Rev.} {\bf D59} (1999) 014017;
\bibitem{JalilianMarian:1997jx}
  J.~Jalilian-Marian, A.~Kovner, A.~Leonidov and H.~Weigert,
  Nucl.\ Phys.\  B {\bf 504}, 415 (1997).
\bibitem{JalilianMarian:1997gr}
  J.~Jalilian-Marian, A.~Kovner, A.~Leonidov and H.~Weigert,
   ``The Wilson renormalization group for low x physics: Towards the high
  Phys.\ Rev.\  D {\bf 59}, 014014 (1999).
\bibitem{JalilianMarian:1997dw}
  J.~Jalilian-Marian, A.~Kovner and H.~Weigert,
   ``The Wilson renormalization group for low x physics: Gluon evolution at
  Phys.\ Rev.\  D {\bf 59}, 014015 (1999).
 \bibitem{Bal}  I. I. Balitsky,   Nucl. Phys. {\bf  B463} (1996) 99;
Phys. Rev. Lett. {\bf 81} (1998) 2024;
              { \it Phys. Rev.} {\bf D60} (1999)  014020;
   Phys. Lett. {\bf B518} (2001) 235;
                  {\em Nucl.Phys.} {\bf A692} (2001) 583.
\bibitem{Kov} Y. V. Kovchegov, {\it Phys.Rev} {\bf D60} (1999) 034008  
\bibitem{Avsar:2006jy}
  E.~Avsar, G.~Gustafson and L.~Lonnblad,
JHEP {\bf 0701}, 012 (2007).
\bibitem{BartKut} J. Bartels, K. Kutak \epjc{53}{2008}521
\bibitem{Kutak:2009zk}  K.~Kutak, Phys.\ Lett.\  B {\bf 675} (2009) 332



\bibitem{Stasto:2000er}
  A.~M.~Stasto, K.~J.~Golec-Biernat and J.~Kwiecinski,
   `
  Phys.\ Rev.\ Lett.\  {\bf 86}, 596 (2001).
s.\ Lett.\  B {\bf 675} (2009) 332[arXiv:0903.3521 [hep-ph]].

\bibitem{GBF}K. Golec-Biernat, M. W\"usthoff, {\it Phys.Rev.} D60:114023,1999
\bibitem{SM2} S. Munier  Xiv:0901.2823
\bibitem{BKphen} K. Golec-Biernat, L. Motyka, A. Stasto, Phys.Rev.D65:074037,2002\\
                 K.Kutak, J. Kwieci\'nski {\it Eur.Phys.J.} C29:521,2003\\
		 K.Kutak, A. Stasto {\it Eur.Phys.J.}C 41:343-351,2005\\
		 K. Kutak, Tsukuba 2006, Deep inelastic scattering* 97-100\\
		 C. Marquet, G. Soyez {\it Nucl.Phys} A760 (2005) 208-222\\
\bibitem{SM1} S. Munier, R. Peschanski{\it Phys.Rev.Lett}.91:232001,2003
\bibitem{MP3} S. Munier R. Peschanski {\it Phys.Rev.} D70:077503,2004
\bibitem{IJM}E. Iancu, K. Itakura, L. Mclerran {\it Nucl.Phys.} A708 (2002) 327-352
\bibitem{enberg} R. Enberg {\it Phys.Rev.}D75:014012,2007

\end{thebibliography}
\end{document}